# Emergent equilibrium and quantum criticality in a two-photon dissipative oscillator

V.Yu. Mylnikov *, S.O. Potashin, G.S. Sokolovskii and N.S. Averkiev

Ioffe Institute, 26 Polytechnicheskaya str., St. Petersburg, 194021, Russia
* Correspondence: vm@mail.ioffe.ru (V.Y.M.);

**Abstract:** We study the dissipative phase transition in a quantum oscillator with two-photon drive and two-photon dissipation. Using the semi-classical Langevin equation and the truncated Wigner approximation, we construct a theory of non-perturbative quantum fluctuations and go beyond the semi-classical approximation. We demonstrate the mapping of a two-photon quantum dissipative oscillator onto a classical equilibrium model of a nonlinear classical oscillator in a white-noise environment. Then, we justify the applicability of the Boltzmann-Gibbs theory for a given dissipative phase transition. To do that, we explicitly demonstrate the Boltzmann-Gibbs-like form of stationary distribution function depending on the effective temperature, which is determined by the two-photon pump rate. In addition, we provide a description of the quantum critical region and obtain critical exponents that appear to be in very good agreement with numerical simulations.

**Introduction**

Both physics and quantum optics communities are currently highly engaged in dissipative phase transition (DPT). When abrupt and nonanalytical changes in the physical observables originate from a direct manipulation of the coupling constants, external driving, or dissipation rates of the system, DPT can be discovered[1,2]. The growing interest in the DPT has its starting point in the rapid expansion of reservoir engineering, a brand-new field of quantum optics. It provides experimental and theoretical studies of dissipative critical phenomena and allows to obtain a wide range of controllable nonequilibrium quantum systems. Recently it was also reported on DPT in circuit quantum electrodynamics waveguides[3–5], systems of atoms[6–8], semiconductor micropillars[9], and nano-optomechanical oscillators[10]. This paper will focus on an optical oscillator with two-photon driving and two-photon dissipation as a prominent example of an open quantum system[11]. A qubit platform based on quantum Schrödinger cat states[12], for utilization in all types of quantum computation[13], is the model's first immediate practical application[14].

Study of numerical techniques typically utilizes for examining nonequilibrium quantum systems. The Liouvillian superoperator's diagonalization[15,16] and the integration of a master equation on a truncated Fock basis[11,17] are the two most noteworthy examples. Numerical study, however, often leaves it challenging to figure out the actual mechanisms underlying the system's behavior. The simplest method of analytical approach that illuminates the physics involved is the semiclassical approximation[18]. It can be successfully applied in a small fluctuation regime far from the critical point. Unfortunately, fluctuations become large in the vicinity of dissipative phase transitions, and semiclassical approximations cease to be applicable[19]. Hence, the only way to advance the analytical description of the DPT, strongly affected by the quantum fluctuations[11], is to develop a nonperturbative theory.

But, nevertheless, the nonequilibrium nature of an open quantum system significantly complicates its consideration beyond the mean-field description. On the other hand, the well-developed Landau theory can be applied to conventional solid-state systems coupled to a thermal reservoir. Expansion of the application field of Landau theory to the nonequilibrium phenomena can greatly simplify the mathematical description of the theory and give new insights to the current understanding of driven-dissipative quantum systems. Recently, mapping from a nonequilibrium quantum problem to equilibrium was successfully demonstrated for the weak-coupling limit of the driven-dissipative Bose-Hubbard model[20], spin models[21], open Dicke model[19], and Kerr oscillator subject to a two-photon driving and one-photon dissipation[22]. For all models, the phase transition between different phases is described by widely known thermodynamic concepts as the thermodynamic potential and the effective temperature. However, finding correspondence between nonequilibrium and thermal problems moved to be challenging being affected by the dissipative and coherent properties of the system.

In this paper, we rigorously demonstrate mapping from the nonequilibrium DPT in a two-photon dissipative oscillator to a nonlinear classical oscillator with noise. To do this, we use the truncated Wigner approximation and derive quasi-classical Langevin equation. This enables us to go beyond the semi-classical approximation and still explicitly take into account critical quantum fluctuations. We justify the well-known statistical physics concepts and establish emergent equilibrium. To do so, we explicitly derive the Boltzmann form of a Wigner function with effective temperature depending on the two-photon pump rate. Justification of the Boltzmann-Gibbs theory to the studied DPT is one of the key results of a present research. Another problem addressed in this paper relates to the fact that fluctuations rapidly grow near the critical point, as in the situation of a thermodynamic phase transition, and the critical region appears where the Landau theory ceases to be applicable. To address this limitation, we explicitly calculate the average number of photons and anomalous average as a function of a two-photon pump rate and the related critical exponent. Thus, the second major outcome of our research is the explanation of quantum criticality, which is strongly affected by the critical quantum fluctuations.

## The model

Let us consider a quantum superconducting microwave cavity (Fig. 1). We shall limit ourselves to the resonant fundamental mode that is described by a quantum oscillator with the Hamiltonian:

$$\hat{H}_0 = \omega_c \hat{a}^\dagger \hat{a}, \qquad (1)$$

where $\hbar=1$, $a/a^\dagger$ are the annihilation/creation bosonic quantum operators, and $\omega_c$ is a fundamental mode frequency. The energy levels of a quantum oscillator are equally-spaced with level spacing equals to the resonant frequency $\omega_c$. The lowest energy state is a vacuum state, which contains no photons. The upper lying state with energy $m\omega_c$ corresponds to $m$ photons in a microwave cavity, where $m$ is a positive integer. In this configuration, the microwave-superconducting oscillator is a closed quantum system. If a system is prepared in a given state, it will remain in that state indefinitely. However, we will consider nonequilibrium generalization of a given system by applying a parametric coherent pump at frequency $\omega_p$, as shown in Fig. 1. The corresponding pump Hamiltonian has the form:

$$\hat{H}_p = \left(G e^{-i2\omega_p t} \hat{a}^{\dagger 2} + G^* e^{i2\omega_p t} \hat{a}^2\right)/2, \qquad (2)$$

where $G$ represents a two-photon pump rate. For the sake of simplicity, we assume that the two-photon pump, $G$, is a real variable. The presence of quadratic creation operators in the Eq. (2) has a clear physical meaning of creating a pair of photons in a cavity by applying a classical microwave tone at a frequency $\omega_p$ per one photon. The physical implementation of this type of squeezing drive can be found in resent work [12]. Let us consider the resonance case at frequency $\omega_c=\omega_p$, when the classical pump creates exactly two-photon pair. The cavity will absorb incoming energy and increase the population of photon pairs over time [23]. In this case, the stationary state is absent and system is highly non-equilibrium.

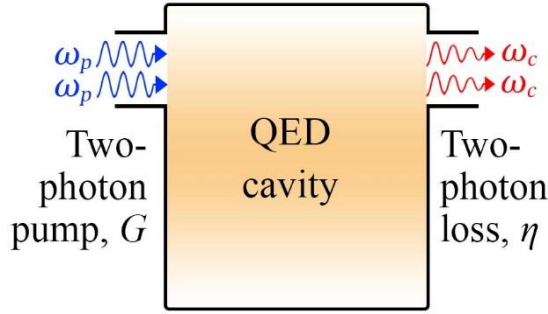

**Figure 1**. A schematic of the photon microwave cavity with two-photon pump at $G$ rate and two-photon dissipation at $\eta$ rate (see text for details).

On opposite side, one can consider the strong dispersive regime far from the resonance ($\omega_p \gg \omega_c$). Let us shift from the laboratory frame to the rotating frame[24], where the resulting Hamiltonian has a time-independent form:

$$\hat{H} = -\Delta \hat{a}^\dagger \hat{a} + G(\hat{a}^{\dagger 2} + \hat{a}^2)/2, \qquad (3)$$

and $\Delta=\omega_p-\omega_c$ denotes a frequency detuning. Far from the resonance, the two-photon drive cannot effectively pump the cavity with photon pairs. Resultingly, energy transfer is suppressed and the stationary state of the system is well defined. However, the presence of a two-photon Hamiltonian changes the ground state from a vacuum to a squeezed vacuum state and alters the energy gap between the ground and first excited states $\sqrt{\Delta^2-G^2}$ [25]. When two-photon pump rate approaches the frequency detuning, energy gap tends to zero. This phenomenon is also known as the "spectral collapse" [26] and it occurs precisely at the exceptional or critical point $\Delta=G$, where the stationary state no longer exists.

Summarizing the above, there are a steady ground state and well-defined excitations at large frequency detuning ($\Delta>G$), but the stationary picture breaks down at the critical point $\Delta=G$. In the small detuning regime ($\Delta<G$), energy is effectively pumped into the cavity, creating two-photon pairs and preventing the stability of the system. However, nonlinearity can stabilize the steady state even for the small detuning parameter area. Further, we consider the nonlinear two-photon dissipation process shown in Fig. 1 and completely determining the behavior of the stationary state below the critical point ($\Delta<G$). The simplest way to introduce dissipation into the quantum system is to couple it to the reservoir. By integrating out all reservoir degrees of freedom, one can derive the effective Lindblad equation on the reduced density matrix [23,27]:

$$\frac{d}{dt}\hat{\rho} = -i\left[\hat{H}, \hat{\rho}\right] + \frac{\eta}{2} D[\hat{a}^2](\hat{\rho}), \qquad (4)$$

where $D[\hat{a}^2](*)=2\hat{a}*\hat{a}^{\dagger 2}-(\hat{a}^{\dagger 2}\hat{a}^2*+*\hat{a}^{\dagger 2}\hat{a}^2)$ is a Liouvillian, and $\eta$ is a two-photon dissipation rate. The quantum dynamic of the system with two-photon pump and two-photon dissipation is entirely determined by Eq. (4). Further, we will investigate the steady state of the system using various approaches, including semi-classical theory, truncated Wigner approximation and Langevin stochastic dynamics.

**Semi-classical theory**

Let us analyze the stationary state when the drive balances the dissipation. As we show further, the quantum fluctuations play a significant role in our problem. However, for weak interactions and large average number of photons, the main properties of the steady state can be understood using the semi-classical approximation [23,24]. Within this approximation, the coherent field amplitude $\alpha(t)=\text{Tr}[\hat{a}\rho(t)]$ obeys the following simple equation:

$$\partial_t \alpha(t) = i\Delta\alpha(t) - iG\alpha^*(t) - \eta\alpha(t)^*\alpha(t)^2. \quad (5)$$

The steady state of the system is characterized by the stationary solutions of the Eq. (5), $\alpha_S = \pm n_S^{1/2} e^{i\phi}$, where average photon number, $n_S$, and the phase factor, $e^{i\phi}$, are given by:

$$n_S = \frac{\sqrt{G^2 - \Delta^2}}{\eta}\theta(g^2 - \Delta^2), \quad \exp[i2\phi] = \frac{-iG}{\sqrt{G^2 - \Delta^2} - i\Delta}, \quad (6)$$

where $\theta(x)$ is the Heaviside step function. We also can find the stationary values of the photonic quadratures. Let us introduce them in the following way [26]:

$$\alpha(t) = (x(t) + ip(t))/\sqrt{2}, \quad (7)$$

where $x(t)=\text{Re}[\alpha(t)]\sqrt{2}$ and $p(t)=\text{Im}[\alpha(t)]\sqrt{2}$. From Eq. (6) one can easily find the stationary semi-classical solution for two photonic quadratures:

$$x_S = \pm\frac{(G+\Delta)^{\frac{3}{4}}(G-\Delta)^{\frac{1}{4}}}{\sqrt{G\eta}}\theta(G^2-\Delta^2), \quad p_S = \mp\frac{(G+\Delta)^{\frac{1}{4}}(G-\Delta)^{\frac{3}{4}}}{\sqrt{G\eta}}\theta(G^2-\Delta^2). \quad (8)$$

Fig. 2 shows the dependence of the average number of photons, $\langle\hat{a}^\dagger\hat{a}\rangle$, and anomalous average, $\langle\hat{a}^2\rangle$, on the frequency detuning. For calculations, we utilize the semi-classical approximation (6) and the P-representation (35) (see methods).

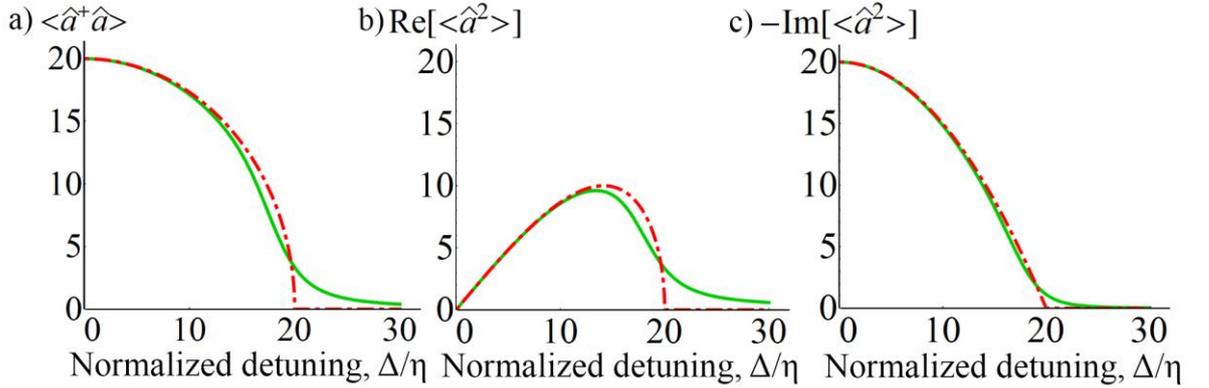

**Figure 2.** (a-c) Average number of photons $n=\langle\hat{a}^\dagger\hat{a}\rangle$ (a), and the real part (b) and minus imaginary part (c) of the anomalous average $\langle\hat{a}^2\rangle$ vs the normalized frequency detuning $\Delta/\eta$. The normalized two-photon pump rate and two-photon dissipation rate are set to $G/\eta=20$ and $\eta=1$, correspondingly. Red dash-dotted line represents semiclassical approximation (6), while green line shows the P-representation (35) (see methods and work of Bartolo et al. [24] for more details).

However, as shown in Fig. 2, the critical region near the critical point cannot be described by the semiclassical theory. As we will demonstrate in the following sections, quantum fluctuations, which was ignored in semiclassical approach, becomes crucial and smear out the sharp dependence of the average photon number typical of the conventional second-order phase transition.

**The truncated Wigner approximation and Langevin equation**

In this section, we will take into account the quantum fluctuations of the system, which significantly affect the DPT near the critical point. The simplest, most attractive, and common method of incorporating quantum fluctuation into our problem is a modification of semi-classical equation (5) by introducing the stochastic noise [20, 22]. To do this correctly, we need to study the Wigner distribution function [28]:

$$W(\alpha,\alpha^*) = \frac{2}{\pi}Tr\left[\hat{D}_\alpha \exp(i\pi\hat{a}^\dagger\hat{a})\hat{D}_\alpha^\dagger \hat{\rho}\right], \quad (9)$$

where $\hat{D}_z=\exp[z\hat{a}^\dagger - z^*\hat{a}]$ is the displacement operator. The Wigner function is a convenient tool for visualizing the properties of the steady state, since it plays the role of a quasi-probability distribution function. It is worth noting that the quantum averaging of any Weyl-ordered products of operators $(...)_S$ [29] can be rewritten in terms of the moments of the Wigner function:

$$\left\langle (\hat{a}^{\dagger m}\hat{a}^{k})_{S}\right\rangle =\left\langle \alpha^{*m}\alpha^{k}\right\rangle =\int \alpha^{*m}\alpha^{k}W(\alpha,\alpha^{*})d^{2}\alpha, \tag{10}$$

where quantum averaging replaces averaging over Wigner function.

By analogy with Eq. (4), one can derive the equation of motion for the Wigner function (see methods). Using truncated Wigner approximation [30,31], in which the third-order derivatives in the exact differential equation are ignored, one can obtain the following Fokker-Planck equation for the Wigner function:

$$\frac{\partial W}{\partial t}=-\left[\frac{\partial}{\partial \alpha}(i\Delta\alpha-iG\alpha^{*}-\eta\alpha^{*}\alpha^{2})+c.c.\right]W+\eta\left[\frac{\partial}{\partial \alpha}\alpha^{*}\frac{\partial}{\partial \alpha^{*}}\alpha+c.c.\right]W. \tag{11}$$

The accurate justification of the truncated approximation is discussed in Methods. The main limitation is a large number of photons in the cavity, i.e. proximity of the system to the "thermodynamic" limit ($\eta \ll g$) [20,32]. It's easy to demonstrate that the Fokker-Planck Eq. (11) is equivalent to the following semi-classical Langevin equation (Methods):

$$\partial_{t}\alpha(t)=i\Delta\alpha(t)-iG\alpha^{*}(t)-\eta\alpha(t)^{*}\alpha(t)^{2}-i\sqrt{2\eta}\alpha(t)^{*}\xi(t), \tag{12}$$

where the Eq. (12) is defined in the Stratonovich sense [33], and $\xi(t)$ is a complex white noise with zero average and the following quadratic correlators:

$$\left\langle \xi(t)\xi(t')\right\rangle =0,\quad \left\langle \xi(t)\xi^{*}(t')\right\rangle =\delta(t-t'). \tag{13}$$

If we neglect the noise in the Langevin equation (12), then it is strictly transformed into the semi-classical equation of motion for the complex field amplitude (5). Interestingly, the noise contribution to the Langevin equation (12) is proportional to the conjugate photonic field amplitude, $\alpha(t)^{*}$. Thus, we are dealing with the Langevin equation governed by multiplicative noise. From a physical point of view, this type of noise occurs as a result of interaction between our system and the reservoir. As discussed above, a pair of photons jumps from the cavity to the reservoir, causing the process of two-photon dissipation. However, an additional process involves the virtual creation and annihilation of a photon pair. Consequently, the two-photon pump rate, $G$, becomes a random variable, $G+(2\eta)^{1/2}\xi(t)$. Therefore, the semi-classical equation of motion (5) becomes the stochastic Langevin equation (12). Here, we use the Wigner distribution to derive the semi-classical Langevin equation (12). However, we believe that the same Langevin equation (12) can be derived using the Keldysh functional integral approach [20].

Now we derive the equation of motion for photonic quadratures (7). The Langevin equation (9) is transformed into the following system of stochastic equations:

$$\partial_{t}x=-(\Delta+G)p-\frac{\eta}{2}x(x^{2}+p^{2})+\sqrt{\eta}(x\xi_{p}-p\xi_{x}),$$
$$\partial_{t}p=(\Delta-G)x-\frac{\eta}{2}p(x^{2}+p^{2})-\sqrt{\eta}(x\xi_{x}+p\xi_{p}), \tag{14}$$

where we rewrite noise as follows: $\xi=(\xi_{x}+i\xi_{p})/\sqrt{2}$. Here the random forces $\xi_{x}$ and $\xi_{p}$ are real variables and have the following correlation functions:

$$\left\langle \xi_{x}(t)\xi_{p}(t')\right\rangle =0,\quad \left\langle \xi_{x}(t)\xi_{x}(t')\right\rangle =\left\langle \xi_{p}(t)\xi_{p}(t')\right\rangle =\delta(t-t'). \tag{15}$$

In the next section we will connect the derived stochastic Langevin equations (14) with the problem of the Brownian particle propagating in the external potential with nonlinear friction. This analogy will help us to find the stationary Wigner distribution function near the critical point.

**The Landau theory and Boltzmann-Gibbs-like stationary distribution function near the critical point**

This section is devoted to the analytical study of the stationary Wigner distribution function. We focus on the parameter area near the critical point, where the properties of the system are significantly affected by quantum fluctuations. The initial strategy was to use the potential solution method[33] to obtain the stationary distribution function of the Fokker-Planck equation (11). However, it is not applicable to our problem because the drift and diffusion coefficients of Eq. (11) do not satisfy the potential conditions. Even so, we can overcome this difficulty by noting that the main simplification follows from the considering the parameter region near the critical point ($|\Delta-G|\ll G$). As shown in Fig. 3a, in such a region, the quadrature variance $<p^2>$ is much smaller than $<x^2>$. Thus, we can neglect the $p$ quadrature with respect to the much more intense $x$ quadrature in the Eqs. (14), and obtain the following system of stochastic equations:

$$\partial_{t}x=-2Gp-\frac{\eta}{2}x^{3},\quad \partial_{t}p=(\Delta-G)x-\frac{\eta}{2}px^{2}-\sqrt{\eta}x\xi_{x}, \tag{16}$$

where we consider only the stochastic force $\propto \xi_{x}$ and neglect noise $\propto \xi_{p}$, because the first term produces a large variation of the normalized $p$-component, while the second term creates the small fluctuations of the normalized $x$-component:

$$\dot{p}/p \propto (x/p)\sqrt{\eta}\xi_x \gg \dot{x}/x \propto \sqrt{\eta}\xi_p. \tag{17}$$

A rigorous justification for the developed system of equations (16) at the critical point can be found in Methods (Scaling at the critical point).

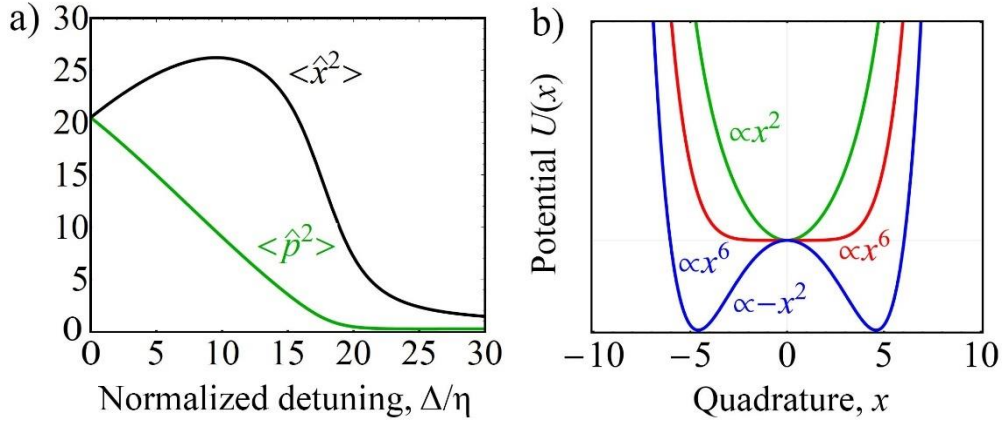

**Figure 3**. (a) The quadrature variances $\langle\hat{x}^2\rangle$ (black line) and $\langle\hat{p}^2\rangle$ (green line) vs the normalized frequency detuning $\Delta/\eta$. (b) The dependence of the effective potential (20) on the quadrature, $x$, for different values of the detuning $\Delta/\eta=17$ (blue line), $\Delta/\eta=20$ (red line), and $\Delta/\eta=23$ (green line). The normalized two-photon pump rate and two-photon dissipation rate are set to $G/\eta=20$ and $\eta=1$.

In the situation of two coupled stochastic equations (16), a general method of fast and slow variables is used to facilitate further calculations [20,22,34]. However, in our case, both quadratures $x$ and $p$ have slow dynamics near the critical point, and the fast-slow separation is not observed in our system. But nevertheless, we can map the problem of the semi-classical nonlinear parametric oscillator (16) onto the problem of the Brownian particle propagating in the external potential with nonlinear friction and noise. Let us introduce a new variable:

$$v(t) = -2Gp(t) - \eta x(t)^3/2. \tag{18}$$

The Langevin Eqs. (16) transform into the following set of equations:

$$\partial_t x = v,$$
$$\partial_t v + 2\eta x^2 v = -\frac{1}{m}\partial_x U(x) + x\sqrt{\frac{4\eta T_{\text{eff}}}{m}}\xi_x, \tag{19}$$

where $2\eta x^2$ is an effective nonlinear damping coefficient, $m=1/2G$, and $U(x)$ is the effective potential:

$$U(x) = \frac{(\Delta-G)}{2}x^2 + \frac{\eta^2}{48G}x^6, \tag{20}$$

which produces the force $-\partial_x U(x)$ in the Eq. (19), and $T_{\text{eff}}$ is an effective [19,20,22,35] or quantum temperature [36], which has the following form:

$$T_{\text{eff}} = G/2. \tag{21}$$

The effective potential (20) is very similar to a previously discussed potential for a Kerr parametric oscillator with one-photon dissipation at both zero [22] and nonzero [34] frequency detuning. However, for the problem of a parametric oscillator with two-photon dissipation and two-photon pump, the effective potential (20) is introduced for the first time. The resulting effective Langevin equation (20) is very similar to the equation of motion of a dumped nonlinear classical oscillator in a white-noise environment. Let us discuss the effective potential (20), which is shown in Fig. 3b and is very similar to the thermodynamic potential in Landau theory. First of all, there is a quadratic term with respect to the photonic field $x$ with the alternating coefficient $(\Delta-G)$ crossing zero at the phase transition point $\Delta=G$. Secondly, this is also a nonlinear term, which ensures the stability of the potential below the critical point ($\Delta<G$) at large values of $x$. It's quite obvious that the quadratic term makes the main contribution to the effective potential above the critical point, when $\Delta>G$, and the nonlinear term can be neglected. However, after passing the critical point, the nonlinear term becomes important when $\Delta<G$. Below the critical point ($\Delta<G$), the semi-classical solution (8) minimizes the effective potential (20), and the Gaussian fluctuations near the semi-classical solution characterize the properties of a new steady state.

The next step is a calculation of the Wigner function using the Langevin Eq. (19). The corresponding Fokker-Planck equation has the form of a Klein–Kramers equation [37]:

$$\partial_t W = \left[ \partial_v \left( \frac{1}{m} \partial_x U(x) + 2\eta x^2 v \right) - \partial_x v + \frac{2\eta x^2}{m} T_{\text{eff}} \partial_v^2 \right] W. \tag{22}$$

We can easily find its stationary solution [33]:

$$W(x,v) = \frac{1}{Z_0} exp\left[ -\frac{mv^2}{2T_{\text{eff}}} - \frac{U(x)}{T_{\text{eff}}} \right], \tag{23}$$

where $Z_0$ is a normalizing constant. The first term in the exponent (23) plays the role of the effective kinetic energy, and the second term relates to the potential energy. It is convenient to return the original photonic quadrature, $p$, using the Eq. (18) and obtain:

$$W(x,p) = \frac{1}{Z_0} exp\left[ -2\left( p + \frac{\eta x^3}{4G} \right)^2 - \frac{U(x)}{T_{\text{eff}}} \right]. \tag{24}$$

One can easily prove that minimizing the exponent argument (24) reproduces the semi-classical theory (8) for both $x$ and $p$ quadratures near the critical point.

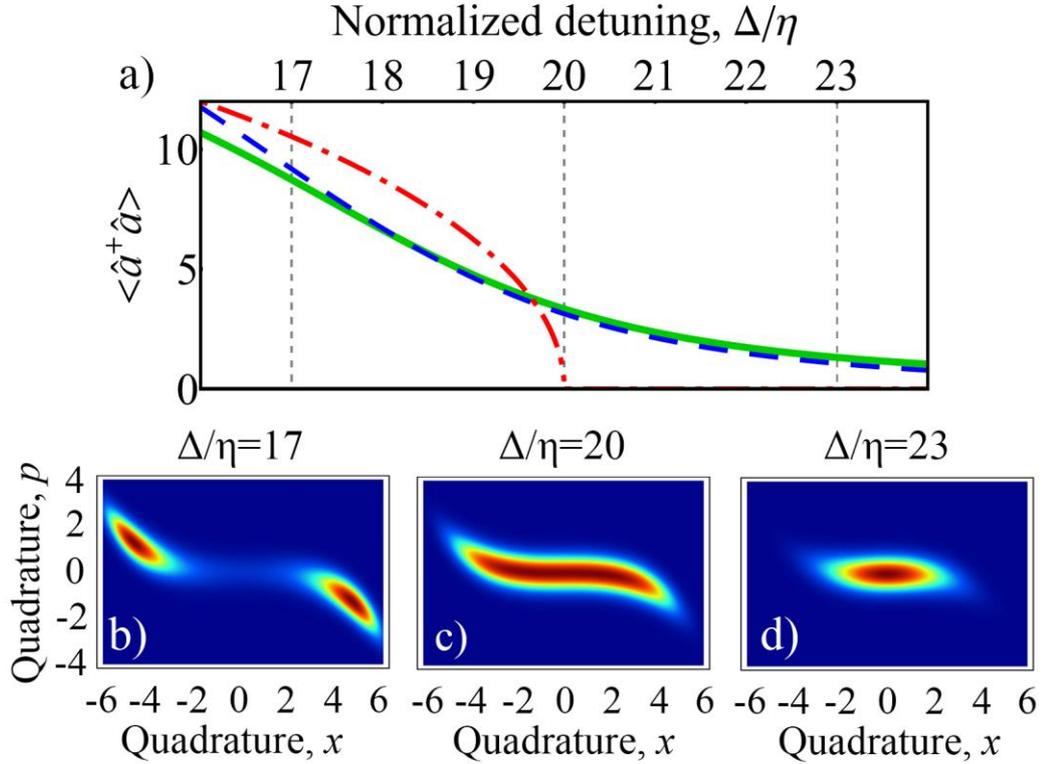

**Figure 4**. (a) Average number of photons $n=<a^\dagger a>$ vs the normalized frequency detuning $\Delta/\eta$. Red dash-dotted line represents semiclassical approximation (6), green line shows the P-representation solution (35), and blue dashed line is obtained using the Boltzmann distribution function (24). (b-d) Steady-state Wigner function $W(x,p)$ calculated using the Boltzmann distribution function (24) for different values of normalized detuning (b) $\Delta/\eta=17$, (c) $\Delta/\eta=20$, and (d) $\Delta/\eta=23$. The normalized two-photon pump rate and two-photon dissipation rate are set to $G/\eta=20$ and $\eta=1$.

As a result, we found a stationary Wigner distribution function (24), valid in the vicinity of the dissipative phase transition and the thermodynamic limit ($G/\eta<<1$). The calculated Wigner function (24) has a Boltzmann-Gibbs-like form with a corresponding Landau potential (20) and effective temperature (21). Thus, we demonstrate the connection between the considered nonequilibrium DPT and the equilibrium thermodynamic phase transition. It's also worth mentioning that we started with a Markovian quantum bath at zero physical temperature. However, the photonic quadrature, $x$, has acquired a new emergent property, manifested in effective thermalisation, also known as the quantum heating phenomenon [35]. Interestingly, the discussed effective temperature (21) is not affected by the dissipation rates, as in the case of the Kerr oscillator with one-photon dissipation [22] or driven-dissipative Bose-Hubbard model [20], but rather by the coherent coupling constant: the two-photon pump rate, $G$, similar to the open Dicke model [19]. It is important that the predicted effective temperature (21) coincides with the temperature calculated in the work of Alex Kamenev et al. [35], devoted to the problem of a quantum parametric oscillator without nonlinearities.

From Fig. 4a it is clear that the emergent thermal fluctuations determined by the effective temperature (21) smear out the dissipative phase transition near the critical point and modify the semi-classical solution (6)-(8). Figs. 4b-d also show Wigner function (24) for different values of the frequency detuning. Below the critical point (Fig. 4b), the Wigner function has two bright peaks associated with two coherent states having opposite signs[13]. At the critical point, two bright peaks merge together (Figs. 4c), and the DPT takes place. Above the critical point ($\Delta>G$), the stationary state transforms into a thermal squeezed vacuum state [35] (Fig. 4d). Consequently, the main characteristics of the DPT are accurately captured by the developed quasi-equilibrium theory. It is worth to mention that the negativity of the Wigner function in Fig. 4b is missing and we observe only a statistical mixture of two coherent states. This happens because derived Langevin equations breaks the strong parity symmetry [16]. Thus, our theory is correct if we assume infinitesimal one-photon dissipation causing decoherence.

Let us introduce the reduced Wigner function:

$$w_R(x) = \int_{-\infty}^{\infty} \frac{dp}{2} W(x,p), \tag{25}$$

where it is normalized in the ordinary sense $\int dx\, w_R(x)=1$. Since the Wigner distribution (24) is quadratic in $p$ quadrature, we can easily calculate the reduced one (25):

$$w_R(x) = \frac{1}{Z} \exp\left[-\frac{U(x)}{T_{\text{eff}}}\right], \tag{26}$$

where $Z = \int dx\, \exp[-U(x)/T_{\text{eff}}]$ is a normalizing constant. Using the reduced Wigner function (26) we can calculate all physical observables of the system in terms of $x$ quadrature:

$$\langle (\hat{p}^k \hat{x}^m)_S \rangle = \langle p^k x^m \rangle = \int_{-\infty}^{\infty} dx \frac{(-1)^k}{(2i\sqrt{2})^k} H_k\left[i\sqrt{2}\frac{\eta}{4G}x^3\right] x^m w_R(x), \tag{27}$$

where $H_k[z]$ is the Hermite polynomials. For example, let us express several moments of the $p$ quadrature through the reduced distribution function using the equation (27):

$$\langle p \rangle = -\frac{\eta}{4G}\langle x^3 \rangle, \quad \langle xp \rangle = -\frac{\eta}{4G}\langle x^4 \rangle, \quad \langle p^2 \rangle = \frac{1}{4} + \frac{\eta^2}{16G^2}\langle x^6 \rangle. \tag{28}$$

The reduced distribution function (26) from Landau theory should be compared with the P-representation (Methods). For this, we need to consider the Wigner function (37) and substitute it into the Eq. (25):

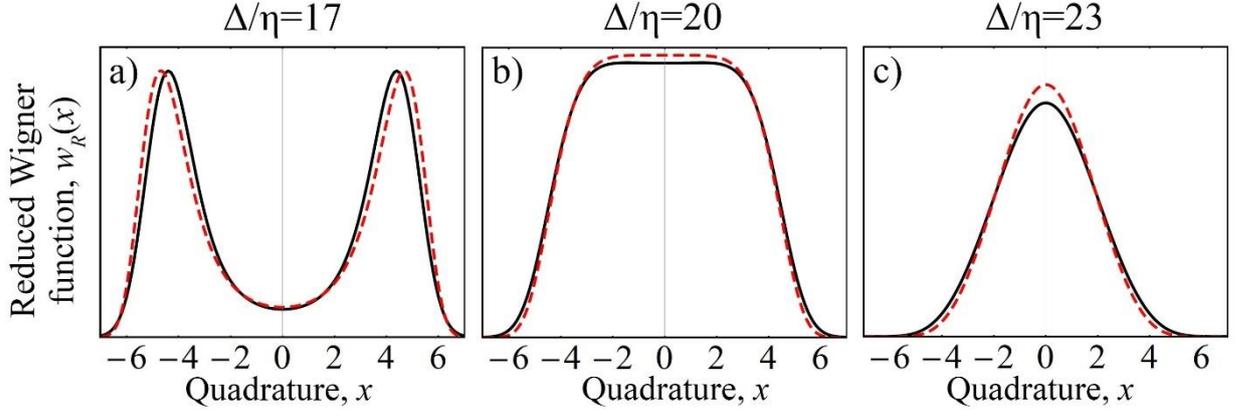

**Figure 5**. (a-c) The reduced Wigner function vs the quadrature, $x$, for different values of detuning $\Delta/\eta=17$ (a), $\Delta/\eta=20$ (b), and $\Delta/\eta=23$ (c). The black line represents the reduced Wigner function, calculated using the full Wigner function (37), and red dashed line is calculated via the Boltzmann-Gibbs theory (26). The normalized two-photon pump rate and two-photon dissipation rate are set to $G/\eta=20$ and $\eta=1$.

The reduced Wigner function, calculated using the Wigner function (37) and the Boltzmann-Gibbs solution (26), is shown in Fig. 5. Thus, it is strikingly clear that the developed Boltzmann-Gibbs theory shows good agreement with the P-representation, highlighting the utility of the present formalism.

**Critical fluctuations at the critical point**

The most interesting phenomena, as we believe, take place near the critical point. Semi-classical theory (8) or minimization of the Landau potential (20) predicts that the quantum observables have a zero value at the critical point. However, we can use the Boltzmann-Gibbs distribution function (24) and exactly calculate the finite quadrature variances for $\Delta=G$:

$$\left\langle \hat{x}^2 \right\rangle = \left\langle x^2 \right\rangle = \frac{\sqrt{\pi}}{3^{2/3}\Gamma(7/6)}\left(\frac{G}{\eta}\right)^{2/3}, \tag{29}$$

$$\left\langle \hat{p}^2 \right\rangle = \left\langle p^2 \right\rangle = \frac{1}{4} + \frac{\eta^2}{16G^2}\left\langle x^6 \right\rangle = \frac{1}{2}, \tag{30}$$

$$\left\langle (\hat{x}\hat{p})_s \right\rangle = \left\langle xp \right\rangle = -\frac{\eta}{4G}\left\langle x^4 \right\rangle = -\frac{3^{2/3}\Gamma(5/6)}{\Gamma(1/6)}\left(\frac{G}{\eta}\right)^{1/3}, \tag{31}$$

where we use the correspondence (27) between the average over the full Wigner function (24) and the reduced one (26), and also use the connection (10) between the problem of operator averaging and averaging over the Wigner function.

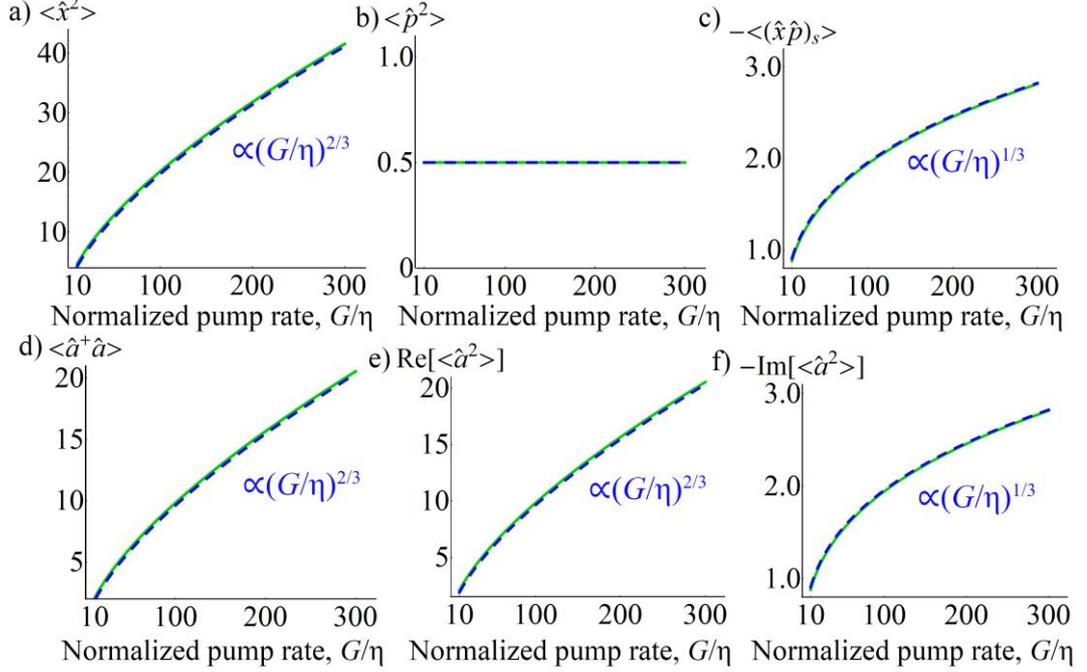

**Figure 6.** (a-c) Dependence of the quadrature variances, $\langle\hat{x}^2\rangle$ and $\langle\hat{p}^2\rangle$, and covariance $-\langle(\hat{x}\hat{p})_s\rangle$ on the normalized two-photon pump rate, $G/\eta$, at the critical point ($\Delta=G$). (d-f) Average number of photons $\langle\hat{a}^+\hat{a}\rangle$, as well as real part and minus imaginary part of the anomalous average $\langle\hat{a}^2\rangle$ vs the normalized two-photon pump rate, $G/\eta$, at the critical point ($\Delta=G$). The green curve is computed using P-representation (35) and (39), and the blue dashed curve is calculated using the Boltzmann-Gibbs theory (29)-(32). The two-photon dissipation rate is set to $\eta=1$.

In a previous section we demonstrated that critical fluctuations have a thermal nature (24). It is fully determined by the finite effective temperature (21), which ensures the thermal population of quantum states by photons and, consequently, a non-zero average number of particles in a stationary state (Fig. 6d). However, Figs. 6a-c and Eqs. (29)-(30) show that the effective thermalisation affects only $x$-quadrature. The second quadrature, $p$, remains finite and still has quantum uncertainty at the critical point (Fig. 6b and Eq. (30)). This fact justifies the previously discussed assumption that $p$-quadrature is much smaller than $x$-quadrature. Nevertheless, the quadratures $x$ and $p$ are highly correlated with each other and therefore have a non-zero covariance (31). Correlations arise because the distribution function (24) cannot be factorized as a function of two independent variables. This is the reason why the bright spot in Figs. 4c does not have the shape of an ellipsoid, as we could see for two independent quadratures, but has a curve producing a non-zero covariance. Worth mentioning is the good agreement of our theoretical calculations with the P-representation. In particular, expressions (29) and (31) correctly predict that the photonic quadrature, $\langle\hat{x}^2\rangle \propto (G/\eta)^{2/3}$, and covariance, $\langle\hat{x}\hat{p}\rangle_s \propto (G/\eta)^{1/3}$, are governed by the power-law with corresponding critical exponents 2/3 and 1/3, correspondingly. Resultingly, quantum fluctuations determine the quantum critical behavior of the system and the corresponding critical scaling of quantum observables.

Now let us focus on the average number of photons $\langle\hat{a}^+\hat{a}\rangle$ and the anomalous average $\langle\hat{a}^2\rangle$ at the critical point. Using the Eq. (39), it is easy to show that both averages can be expressed in terms of the previously calculated quadrature variances (29)-(31). As a result, the behavior of the average number of photons and the anomalous average at the critical point is determined as follows:

$$\langle\hat{a}^{\dagger}\hat{a}\rangle=\frac{1}{2}\langle\hat{x}^2+\hat{p}^2-1\rangle\propto(G/\eta)^{2/3},\ \text{Re}\langle\hat{a}^2\rangle=\frac{1}{2}\langle\hat{x}^2-\hat{p}^2\rangle\propto(G/\eta)^{2/3},\ \text{Im}\langle\hat{a}^2\rangle=\langle(\hat{x}\hat{p})_s\rangle\propto-(G/\eta)^{1/3}. \quad (32)$$

Note that the critical exponents of the average number of photons $\langle\hat{a}^{\dagger}\hat{a}\rangle$ and the real part of the anomalous average Re$[\langle\hat{a}^2\rangle]$ coincide due to the smallness of the $p$-quadrature relative to the $x$-quadrature. However, the imaginary part of the anomalous average has a different critical exponent, which is determined by covariance (31). The average number of photons, as well as the real and imaginary parts of the anomalous average, are shown in Figs. 6d-f. One can see a good agreement between the developed effective equilibrium theory and the P-representation.

To conclude this section, behavior of the higher order operators should be discussed. Let us introduce the normalized second-order correlation function as follows [23]:

$$g^{(2)}=\frac{\langle\hat{a}^{\dagger 2}\hat{a}^2\rangle}{\langle\hat{a}^{\dagger}\hat{a}\rangle^2}=\frac{\langle x^4\rangle+2\langle x^2p^2\rangle+\langle p^4\rangle-4\langle x^2\rangle-4\langle p^2\rangle+2}{(\langle x^2\rangle+\langle p^2\rangle-1)^2}, \quad (33)$$

where we also express the operator problem through the averaging over Wigner function. Analysis of the second-order correlation function provides information about the statistical properties of light in the cavity. Fig. 7 shows the dependence of the normalized second-order correlation function (33) on the frequency detuning.

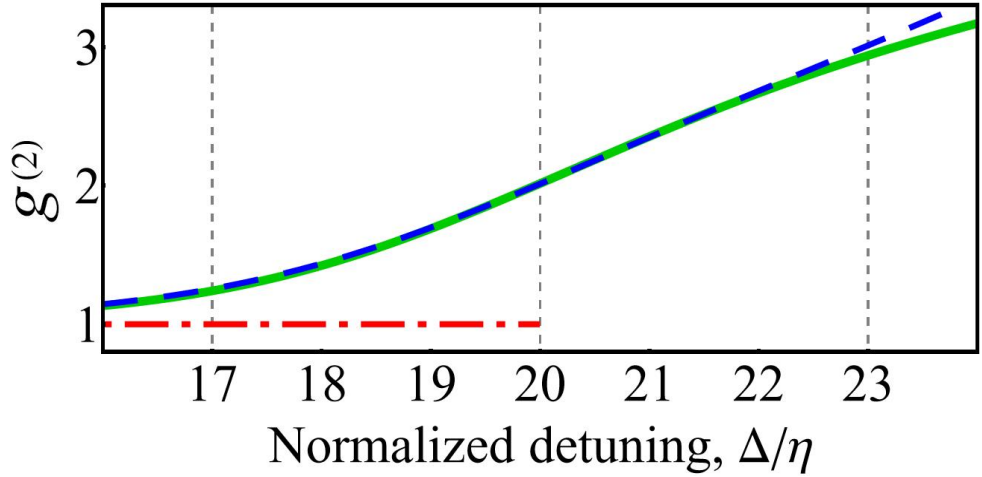

**Figure 7**. The normalized second-order correlation function, $g^{(2)}$, as a function of the normalized detuning, $\Delta/\eta$. Red dash-dotted line represents semiclassical approximation (6), green line shows the P-representation (35), and blue dashed line is obtained using the Boltzmann distribution function (24). The normalized two-photon pump rate and two-photon dissipation rate are set to $G/\eta=20$ and $\eta=1$.

One can see that the correlation function, $g^{(2)}$, below the critical point ($\Delta<G$) is approximately equal to unity. This happened because the system goes into the semi-classical regime, and the correlators are decoupled in the following sense: $g^{(2)}=\langle a^{\dagger 2}a^2\rangle/\langle a^{\dagger}a\rangle^2\approx|\alpha_S|^4/|\alpha_S|^4=1$, where $\alpha_S$ is a semiclassical complex field amplitude. Thus, light in the cavity is coherent below the phase transition point. Above the critical point ($\Delta>G$), the correlation function, $g^{(2)}$, approaches the value of three. This fact can be easily explained using the simple Wick theorem. According to it, the forth-order correlator can be decoupled as follows $\langle a^{\dagger 2}a^2\rangle\approx 2\langle a^{\dagger}a\rangle+\langle a^2\rangle$ [11]. Above the critical point, the anomalous average is approximately equal to the average photon number, and the correlation function, $g^{(2)}$, will be equal to three. At the critical point ($\Delta=G$), we can calculate the correlation function (33) in the large average photon number regime, using the assumption that $x>>p$, and obtain the following simple relation:

$$g^{(2)}\approx\langle x^4\rangle/\langle x^2\rangle^2=2, \quad (34)$$

where averaging over reduced Wigner function (26) is utilized. When the correlation function is equal to two, the light in the cavity is bunched and has a thermal statistic. This fact confirms the previously discussed statement that the emergent equilibrium takes place at the critical point.

**Conclusions and Outlook**

In this paper, we investigate the dissipative phase transition near the critical point arising in a quantum oscillator with two-photon drive and two-photon dissipation. Quantum fluctuations are studied using the Langevin equation and the truncated Wigner approximation. All this allowed us to go beyond the semi-classical approximation and find the Wigner distribution function near the critical point. We demonstrate how the well-known concepts from statistical physics arise in the considered nonequilibrium system and produce emergent equilibrium. To do this, we explicitly demonstrate that the Wigner distribution function has the form of a Boltzmann-Gibbs distribution depending on the effective temperature, which is determined by the two-photon pump rate. This is one of the main results of this work, which justifies the concept of free energy and the applicability of the Landau theory for the considered dissipative phase transition. We also develop the theory of dissipative phase transition at the critical point in the thermodynamic limit, where quantum fluctuations become large and Landau theory ceases to be applicable. Using the obtained Boltzmann-Gibbs-like distribution function, we predict the critical scaling, which is in good agreement with P-representation. This description of quantum criticality is the second main result of our work. We hope that the results of our study will help to look at the problems of open quantum systems through the prism of statistical physics and the theory of thermodynamic phase transitions. Currently, our model has an experimental realization discussed in paper of Leghtas et al. and Berdou et al. [3,12]. It can also stimulate further experimental progress in the field of quantum optics associated with superconducting QED systems. As we believe, the developed theory can be beneficial in building the error corrected logical qubits [38].

**Acknowledgements**

G.S.S. thanks the Russian Science Foundation (project 21-72-30020) for financial support. We acknowledge fruitful discussions with Valentin Kachorovskii, Ksenia Mylnikova, Danil Bugakov, Charles Andrew Downing, Sebastian Diehl and Alex Kamenev.

**Author Contributions**

**Competing interests**

The author declares no competing interests.

**Additional information**

Correspondence and requests for materials should be addressed to V.Yu. Mylnikov.

**Data availability**

Data underlying the results presented in this paper are available from the corresponding authors upon reasonable request.

**Methods**

**P-representation formalism**

For the problem of a two-photon driven nonlinear quantum resonator with pair dissipation, exact solutions can be obtained via the complex P-representation formalism [24]. We focus our attention on the average photon number, anomalous average and the fourth order correlator, which have the following form [24]:

$$\langle \hat{a}^\dagger \hat{a} \rangle = \frac{2|G|^2}{4\Delta^2 + \eta^2} \frac{{}_1F_2\left(\frac{3}{2}; \frac{3}{2} - \frac{i\Delta}{\eta}, \frac{3}{2} + \frac{i\Delta}{\eta}; \frac{|G|^2}{\eta^2}\right)}{{}_1F_2\left(\frac{1}{2}; \frac{1}{2} - \frac{i\Delta}{\eta}, \frac{1}{2} + \frac{i\Delta}{\eta}; \frac{|G|^2}{\eta^2}\right)}, \quad \langle \hat{a}^2 \rangle = \frac{-G}{2\Delta + i\eta} \frac{{}_1F_2\left(\frac{3}{2}; \frac{3}{2} - \frac{i\Delta}{\eta}, \frac{1}{2} + \frac{i\Delta}{\eta}; \frac{|G|^2}{\eta^2}\right)}{{}_1F_2\left(\frac{1}{2}; \frac{1}{2} - \frac{i\Delta}{\eta}, \frac{1}{2} + \frac{i\Delta}{\eta}; \frac{|G|^2}{\eta^2}\right)}, \quad (35)$$

$$\langle \hat{a}^{\dagger 2} \hat{a}^2 \rangle = \frac{|G|^2}{4\Delta^2 + \eta^2} \frac{{}_2F_3\left(\frac{3}{2}, \frac{3}{2}; \frac{1}{2}, \frac{3}{2} - \frac{i\Delta}{\eta}, \frac{3}{2} + \frac{i\Delta}{\eta}; \frac{|G|^2}{\eta^2}\right)}{{}_1F_2\left(\frac{1}{2}; \frac{1}{2} - \frac{i\Delta}{\eta}, \frac{1}{2} + \frac{i\Delta}{\eta}; \frac{|G|^2}{\eta^2}\right)}, \quad (36)$$

where ${}_pF_q(a_1,\ldots,a_p; b_1,\ldots,b_q; z)$ denotes a generalized hypergeometric function [39]. The Wigner function in discussed formalism is given by:

$$W^{(ex)}(x,p) = \frac{2}{\pi} \frac{\left|{}_0F_1\left(\frac{1}{2} - \frac{i\Delta}{\eta}; -i\frac{G}{2\eta}(x-ip)^2\right)\right|^2}{{}_1F_2\left(\frac{1}{2}; \frac{1}{2} - \frac{i\Delta}{\eta}, \frac{1}{2} + \frac{i\Delta}{\eta}; \frac{|G|^2}{\eta^2}\right)} e^{-(x^2+p^2)}, \quad (37)$$

where we use definition (7) that expresses annihilation operator, $a$, in terms of the real-valued quadratures $x$ and $p$. The quantum observables can be calculated from the Wigner function (37) as follows:

$$\left\langle (\hat{x}^k \hat{p}^m)_S \right\rangle = \int_{-\infty}^{\infty} \int_{-\infty}^{\infty} \frac{dxdp}{2} W^{(ex)}(x,p) x^k p^m, \quad (38)$$

where $(…)_S$ symmetrizes or Weyl-orders products of operators[29]. For simplicity, the quadrature variance can be expressed in terms of the average number of photons and anomalous average (35):

$$\langle \hat{x}^2 \rangle = \langle (\hat{a}+\hat{a}^\dagger)^2 \rangle / 2 = \langle \hat{a}^\dagger \hat{a} \rangle + \text{Re}[\langle \hat{a}^2 \rangle] + 1/2,$$
$$\langle \hat{p}^2 \rangle = -\langle (\hat{a}-\hat{a}^\dagger)^2 \rangle / 2 = \langle \hat{a}^\dagger \hat{a} \rangle - \text{Re}[\langle \hat{a}^2 \rangle] + 1/2, \quad (39)$$
$$\langle (\hat{x}\hat{p})_S \rangle = \frac{1}{2}(\langle \hat{x}\hat{p} \rangle + \langle \hat{p}\hat{x} \rangle) = \text{Im}[\langle \hat{a}^2 \rangle].$$

**Truncated Wigner approximation**

We use the truncated Wigner approximation to describe quantum fluctuations. Let us write the equation of motion for the Wigner distribution function using the Eq. (4) on the density matrix and the following correspondence rules [23]:

$$\hat{a}\hat{\rho} \to \left(\alpha + \frac{1}{2}\frac{\partial}{\partial \alpha^*}\right)W[\alpha,\alpha^*], \quad \hat{\rho}\hat{a} \to \left(\alpha - \frac{1}{2}\frac{\partial}{\partial \alpha^*}\right)W[\alpha,\alpha^*],$$
$$\hat{a}^\dagger \hat{\rho} \to \left(\alpha^* - \frac{1}{2}\frac{\partial}{\partial \alpha}\right)W[\alpha,\alpha^*], \quad \hat{\rho}\hat{a}^\dagger \to \left(\alpha^* + \frac{1}{2}\frac{\partial}{\partial \alpha}\right)W[\alpha,\alpha^*]. \quad (40)$$

Applying the transformation rules (40) to the Lindblad Eq. (4), we obtain:

$$\frac{\partial W}{\partial t} = -\left[\frac{\partial}{\partial \alpha}(i\Delta\alpha - iG\alpha^* - \eta\alpha(\alpha^*\alpha - 1)) + c.c.\right]W[\alpha,\alpha^*] +$$
$$+ 2\eta\frac{\partial^2}{\partial\alpha\partial\alpha^*}\left(\alpha^*\alpha - \frac{1}{2}\right)W[\alpha,\alpha^*] + \frac{\eta}{4}\left(\frac{\partial^3}{\partial\alpha\partial(\alpha^*)^2}\alpha^* + c.c.\right)W[\alpha,\alpha^*]. \quad (41)$$

Next step is a justification of the truncated Wigner approximation. Let us introduce the "thermodynamic" limit for our problem [20]:

$$\eta = \tilde{\eta}/N, \alpha = \sqrt{N}\tilde{\alpha}, \partial_\alpha = \partial_{\tilde{\alpha}}/\sqrt{N}. \quad (42)$$

In the "thermodynamic" limit ($N \to \infty$), nonlinearity goes to zero ($\eta \to 0$), average photon number goes to infinity ($<\alpha^*\alpha> \to \infty$), but in Eq. (41) contributions to the drift term associated with the semi-classical theory (5) stay finite:

$$-\left[\frac{\partial}{\partial\alpha}(i\Delta\alpha - iG\alpha^* - \eta\alpha\alpha^*\alpha) + c.c.\right]W[\alpha,\alpha^*] = -\left[\frac{\partial}{\partial\tilde{\alpha}}(i\Delta\tilde{\alpha} - iG\tilde{\alpha}^* - \tilde{\eta}\tilde{\alpha}\tilde{\alpha}^*\tilde{\alpha}) + c.c.\right]W[\tilde{\alpha},\tilde{\alpha}^*]. \quad (43)$$

All other terms in Eq. (41) are small and can be neglected. Thus, the quantum fluctuations don't play a significant role and the semi-classical theory is correct in the "thermodynamic" limit. However, let us stay in the vicinity of the "thermodynamic" limit ($N \gg 1$), where the physical nonlinearity is small ($\eta = \tilde{\eta}/N \ll G$), but has a finite value. The Eq. (41) transforms as given:

$$\frac{\partial W}{\partial t} = -\left[\frac{\partial}{\partial \tilde{\alpha}}(i\Delta\tilde{\alpha} - iG\tilde{\alpha}^* - \tilde{\eta}\tilde{\alpha}(\tilde{\alpha}^*\tilde{\alpha} - \frac{1}{N})) + c.c.\right]W[\tilde{\alpha},\tilde{\alpha}^*] +$$
$$+ 2\tilde{\eta}\frac{\partial^2}{\partial\tilde{\alpha}\partial\tilde{\alpha}^*}\left(\frac{\tilde{\alpha}^*\tilde{\alpha}}{N} - \frac{1}{2N^2}\right)W[\tilde{\alpha},\tilde{\alpha}^*] + \frac{\tilde{\eta}}{4N^2}\left(\frac{\partial^3}{\partial\tilde{\alpha}\partial(\tilde{\alpha}^*)^2}\tilde{\alpha}^* + c.c.\right)W[\tilde{\alpha},\tilde{\alpha}^*]. \quad (44)$$

To take into account the quantum fluctuations, it is necessary to modify the semi-classical theory and consider next-order corrections proportional to $1/N$. Thus, in Eq. (44) we can neglect contributions proportional to $1/N^2$. As a result, we get:

$$\frac{\partial W}{\partial t} = -\left[\frac{\partial}{\partial\alpha}(i\Delta\alpha - iG\alpha^* - \eta\alpha(\alpha^*\alpha - 1)) + c.c.\right]W[\alpha,\alpha^*] + 2\eta\frac{\partial^2}{\partial\alpha\partial\alpha^*}\alpha^*\alpha W[\alpha,\alpha^*], \quad (45)$$

where we return to the original variables (42). The resulting equation (45) takes the form of the Fokker-Planck equation, which corresponds to the following Langevin equation in the Ito sense [33]:

$$\frac{\partial \alpha}{\partial t} = i\Delta\alpha - iG\alpha^* - \eta\alpha(\alpha^*\alpha - 1) - i\sqrt{2\eta}\alpha^*\xi, \quad (46)$$

where $\xi(t)$ is a white noise with correlation functions (13). One can see from Eq. (46) that the quantum fluctuations modify the semi-classical Eq. (5) in two ways. The first term is the random force $\propto \xi(t)$. The second modification is an additional new term, $\eta\alpha(t)$, in Eq. (46). As we will show below, the second contribution can be eliminated using the transformation of the Ito stochastic differential equation to Stratonovich form.

**Connection between the Ito and Stratonovich form of the Langevin equation**

The Langevin equation (46) is written in the Ito sense. For the sake of simplicity, let us transform it to the Stratonovich form, using the standard procedure [33]. Equation (46) can be represented as follows:

$$\frac{\partial \vec{\psi}}{\partial t} = \vec{A} + \hat{B}\vec{\xi}, \tag{47}$$

where $\vec{\psi} = (\psi_1, \psi_2) = (\alpha, \alpha^*)$, $\vec{A}$ is a drift vector:

$$\vec{A} = \begin{pmatrix} i\Delta\alpha - iG\alpha^* - \eta\alpha(\alpha^*\alpha - 1) \\ -i\Delta\alpha^* + iG\alpha - \eta\alpha^*(\alpha^*\alpha - 1) \end{pmatrix} = \begin{pmatrix} i\Delta\psi_1 - iG\psi_2 - \eta\psi_1(\psi_2\psi_1 - 1) \\ -i\Delta\psi_2 + iG\psi_1 - \eta\psi_2(\psi_2\psi_1 - 1) \end{pmatrix}, \tag{48}$$

$\hat{B}$ is a diffusion matrix:

$$\hat{B} = \sqrt{\eta}\begin{pmatrix} -i\alpha^* & \alpha^* \\ i\alpha & \alpha \end{pmatrix} = \sqrt{\eta}\begin{pmatrix} -i\psi_2 & \psi_2 \\ i\psi_1 & \psi_1 \end{pmatrix}, \tag{49}$$

and $\vec{\xi}$ is a noise:

$$\vec{\xi} = \sqrt{2}(\text{Re}[\xi], \text{Im}[\xi]). \tag{50}$$

Using the well-known correspondence between Ito and Stratonovich formulation [33], one can obtain:

$$A_i^S = A_i - \frac{1}{2}\sum_{j,k=1,2} B_{k,j}\frac{\partial}{\partial \psi_k}B_{i,j}, \tag{51}$$

$$B_{i,j}^S = B_{i,j}, \tag{52}$$

where $A_i^S$ and $B_{i,j}^S$ are the drift and diffusion coefficient in the Stratonovich formulation. Substituting expression (49) into (51), we get:

$$A_1^S = A_1 - \eta\psi_1 = i\Delta\psi_1 - iG\psi_2 - \eta\psi_2\psi_1^2 = i\Delta\alpha - iG\alpha^* - \eta\alpha^*\alpha^2,$$
$$A_2^S = A_2 - \eta\psi_2 = -i\Delta\psi_2 + iG\psi_1 - \eta\psi_1\psi_2^2 = -i\Delta\alpha^* + iG\alpha - \eta\alpha\alpha^{*2}, \tag{53}$$

As a result, the Langevin equation in the Stratonovich representation has the form (12).

**Scaling at the critical point**

In this section we will discuss the Langevin equation (14) at the critical point ($\Delta = G$). We will show that equation (16) can be rigorously justified using the scaling formalism. First, consider the Langevin equation for the quadratures $x$ and $p$ at the critical point:

$$\partial_t x = -2Gp - \frac{\eta}{2}x(x^2 + p^2) + \sqrt{\eta}(x\xi_p - p\xi_x),$$
$$\partial_t p = -\frac{\eta}{2}p(x^2 + p^2) - \sqrt{\eta}(x\xi_x + p\xi_p). \tag{54}$$

Next, we scale the two-photon dissipation rate $\eta$, as well as the quadratures $x$ and $p$ and time $t$, by introducing the parameter $N$ as follows:

$$\eta \to \eta/N, \quad x \to N^\nu x, \quad p \to N^\mu p, \quad t \to N^\varepsilon t, \tag{55}$$

where $\nu$, $\mu$ and $\varepsilon$ are the critical exponents, describing the behavior of system at the critical point. As a result of transformation (55), the Langevin equations (54) will have the following form:

$$\partial_t x = -2GpN^{\varepsilon+\mu-\nu} - \frac{\eta}{2}x(x^2 N^{\varepsilon+2\nu-1} + p^2 N^{\varepsilon+2\nu-1}) + \sqrt{\eta}\left(x\xi_p N^{\frac{\varepsilon-1}{2}} - p\xi_x N^{\frac{\varepsilon-1}{2}+\mu-\nu}\right), \tag{56}$$

$$\partial_t p = -\frac{\eta}{2}p(x^2 N^{\varepsilon+2\nu-1} + p^2 N^{\varepsilon+2\mu-1}) - \sqrt{\eta}\left(x\xi_x N^{\frac{\varepsilon-1}{2}-\mu+\nu} + p\xi_p N^{\frac{\varepsilon-1}{2}}\right), \tag{57}$$

As shown below, the Eqs. (56)-(57) contain only one contribution proportional to the two-photon pump rate, $G$. This contribution will be relevant if the following condition is met:

$$\varepsilon = \nu - \mu. \tag{58}$$

Using the Eq. (58), we can express (56)-(57) in terms of remaining critical exponents:

$$\partial_t x = -2Gp - \frac{\eta}{2} x(x^2 N^{3\nu-\mu-1} + p^2 N^{\nu+\mu-1}) + \sqrt{\eta}\left(x\xi_p N^{\frac{\nu-\mu}{2}-\frac{1}{2}} - p\xi_x N^{\frac{\nu-\mu}{2}-\frac{1}{2}}\right), \tag{59}$$

$$\partial_t p = -\frac{\eta}{2} p(x^2 N^{3\nu-\mu-1} + p^2 N^{\nu+\mu-1}) - \sqrt{\eta}\left(x\xi_x N^{3\frac{\nu-\mu}{2}-\frac{1}{2}} + p\xi_p N^{\frac{\nu-\mu}{2}-\frac{1}{2}}\right). \tag{60}$$

Now suppose that the critical slowing down in time dynamic occurs at the critical point. As long as nonlinearity gets smaller, the typical time scale becomes larger. Thus, the critical exponent, $\varepsilon$, should be positive. Using this fact and Eq. (58), it can be argued that the $x$-quadrature at the critical point is much more intense than the $p$-quadrature, since $\nu>\mu$. With all these criteria met, the most relevant terms from the nonlinear sector in Eqs. (59)-(60) will be proportional to $N^{3\nu-\mu-1}$, and from the noise sector will be proportional to $N^{3(\nu-\mu)/2-1/2}$. If we want them to be scale-invariant, the following conditions need to be satisfied:

$$\nu = 1/3, \quad \mu = 0, \quad \varepsilon = 1/3. \tag{61}$$

It is important to note that the critical exponents (61) coincide with the Eqs. (29)-(31) (see section Critical fluctuations at the critical point). Substituting (61) into Eqs. (59)-(60), one can obtain:

$$\partial_t x = -2Gp - \frac{\eta}{2} x(x^2 + p^2 N^{-2/3}) + \sqrt{\eta}\left(x\xi_p N^{-1/3} - p\xi_x N^{-2/3}\right), \tag{62}$$

$$\partial_t p = -\frac{\eta}{2} p(x^2 + p^2 N^{-2/3}) - \sqrt{\eta}\left(x\xi_x + p\xi_p N^{-1/3}\right). \tag{63}$$

It is easy to see that near the thermodynamic limit ($N\gg 1$) the equation (16) is valid, since contributions proportional to $xp^2$ in eq. (62) and $p^3$ in Eq. (63) decrease as $N^{-2/3}$. Of all the noise contributions, a term proportional to $x\xi_x$ remains, since the rest terms also decrease as $N^{-1/3}$ or $N^{-2/3}$.